# Heart Rate Monitoring During Different Lung Volume Phases Using Seismocardiography

Amirtaha Taebi[1,*], *Student Member*, *IEEE*, Andrew J Bomar[2], Richard H Sandler[1,2], Hansen A Mansy[1], *Member*, *IEEE*
[1]Biomedical Acoustics Research Laboratory, University of Central Florida, Orlando, FL 32816, USA
[2]College of Medicine, University of Central Florida, Orlando, FL 32827, USA

{taebi@knights., bomaraj@knights., hansen.mansy@} ucf.edu, richard.sandler@nemours.org

*Abstract*—Seismocardiography (SCG) is a non-invasive method that can be used for cardiac activity monitoring. This paper presents a new electrocardiogram (ECG) independent approach for estimating heart rate (HR) during low and high lung volume (LLV and HLV, respectively) phases using SCG signals. In this study, SCG, ECG, and respiratory flow rate (RFR) signals were measured simultaneously in 7 healthy subjects. The lung volume information was calculated from the RFR and was used to group the SCG events into low and high lung-volume groups. LLV and HLV SCG events were then used to estimate the subjects HR as well as the HR during LLV and HLV in 3 different postural positions, namely supine, 45 degree heads-up, and sitting. The performance of the proposed algorithm was tested against the standard ECG measurements. Results showed that the HR estimations from the SCG and ECG signals were in a good agreement (bias of 0.08 bpm). All subjects were found to have a higher HR during HLV ($HR_{HLV}$) compared to LLV ($HR_{LLV}$) at all postural positions. The $HR_{HLV}/HR_{LLV}$ ratio was 1.11±0.07, 1.08±0.05, 1.09±0.04, and 1.09±0.04 (mean±SD) for supine, 45 degree-first trial, 45 degree-second trial, and sitting positions, respectively. This heart rate variability may be due, at least in part, to the well-known respiratory sinus arrhythmia. HR monitoring from SCG signals might be used in different clinical applications including wearable cardiac monitoring systems.

I. INTRODUCTION

Heart rate (HR) is a common parameter for monitoring cardiovascular function and can identify some abnormalities such as arrhythmia. There are various HR estimation methods that are mostly based on ECG signal processing.

Seismocardiography (SCG) is a technique that measures heart induced vibrations at the chest surface [1]. SCG signal is similar to the familiar phonocardiography, but is largely in the subsonic (below 20 Hz) range. As such, SCG-based methods might provide information that are complementary to other cardiac monitoring methods such as electrocardiography and echocardiography [2]–[5]. SCG signals can also be used for HR estimation. For example, Cosoli *et al.* [6] suggested a general algorithm that can estimate HR from various signals, including SCG, ECG, phonocardiogram (PCG), and photoplethysmogram (PPG). Considering the ECG signal as a gold standard in their study, the SCG HR estimation was more accurate than PCG and PPG. Wahlstrom *et al.* [7] used a Hidden Markov Model to determine different stages of a cardiac cycle which were then used for estimating beat-to-beat intervals. HR and HR variability can be estimated from beat-to-beat intervals of an SCG signal. Mafi [8] suggested an algorithm based on empirical mode decomposition and empirical wavelet transform that can extract HR from SCG signals. Tadi *et al.* [9] used a Hilbert adaptive beat identification technique to determine the heartbeat timings and inter-beat time intervals from SCG signals. An android application was implemented based on this algorithm that can monitor the subjects' heart rate in real time using accelerometers [9]. Tadi *et al.* [10] proposed an algorithm based on S-transform, Shannon energy, and successive mean quantization transform to identify heartbeat and beat-to-beat interval from SCGs. The latter two algorithms had a high agreement with the ECG inter-beat interval values.

Previous studies reported that the SCG signals can have different morphology during high and low lung volume phases [11]–[14]. This study presents an algorithm for heart rate monitoring using seismocardiography during low and high lung volumes. Materials and methods are described in section II. Results are presented and discussed in sections III and IV, respectively, followed by conclusions in section V.

II. METHODOLOGY

*A. Participants*

The study protocol was approved by the Institutional Review Board of the University of Central Florida, Orlando, FL. A total of 7 young male subjects were included in our study. The subjects gave their informed consent, and reported no history of cardiovascular disease. The age, height, and weight of the subjects are reported in Table I.

*B. Human Studies*

For consistency in breathing pattern and tidal volume, the subjects were first trained to control their inspiratory:expiratory

This study was supported by NIH R44HL099053.

(I:E) ratio and respiratory rate utilizing a volume controlled ventilator (Model: 613, Harvard Apparatus, South Natick, MA). The I:E and respiratory rate were set to 1:3 and 12 breath per minute, respectively. For each subject, the tidal volume (TV) was calculated in real time from the respiratory flow rate signal. The TV was displayed to the subjects on a computer screen during the experiment and was kept at 10 to 15 mL/kg.

The subjects were asked to rest on a folding bed and the signals of interest were recorded in three different postural positions; 90 degree (sitting), 45 degree, and 0 degree (supine). The signals were recorded for 1.5 minute at 0 and 90 degrees, while they were acquired for a longer time period (2 trials of 5 minutes each) at 45 degree. The longer time was to check for signal drift over time.

*C. Instrumentation*

A triaxial accelerometer (Model: 356A32, PCB Piezotronics, Depew, NY) was used to record all SCG signals. A signal conditioner (Model: 482C, PCB Piezotronics, Depew, NY) was then used to amplify the accelerometer output with a gain factor of 100. The sensor was affixed to the left sternal border at the level of the 4[th] intercostal space using a double-sided medical-grade tape. This location was chosen due to its high signal-to-noise ratio [15], [16]. The accelerometer's z-axis was aligned perpendicular to the subject chest surface, while the x- and y-axes were aligned parallel to the axial and mediolateral directions, respectively. The respiratory flow rate of the subjects was measured using a pre-calibrated spirometer (Model: A-FH-300, iWorx Systems, Inc., Dover, NH). The inspiration and expiration produced positive and negative flow rate signal amplitudes, respectively. The voltage for respiratory flow rate, ECG, and SCG signals were acquired simultaneously using a Control Module (Model: IX-TA-220, iWorx Systems, Inc., Dover, NH). The lung volume was calculated as the integral of the respiratory flow rate.

All the signals were acquired simultaneously at a sampling frequency of 10 kHz and down-sampled to 320 Hz. The SCG signals were then filtered using a low-pass filter with a cut-off of 100 Hz to remove the remaining respiratory sound noise, which have significant energy above this cut-off frequency [17]. Matlab (R2015b, The MathWorks, Inc., Natick, MA) was used to process all signals.

*D. Heart Rate Monitoring*

Previous studies [11], [12] suggested that the lung volume can affect the SCG signal morphology. In this study, after SCG signal segmentation, the lung volume signal was used to group the SCG cycles into two groups of high and low lung volume (HLV and LLV, respectively). Fig. 1 shows a 10-seconds portion of a SCG signal lined up with the lung volume. The LLV and HLV parts of the lung volume were labeled in the figure. The SCG events that occurred during low and high lung volumes were subsequently called LLV and HLV SCG events, respectively. These events were then used to measure the heart rate during LLV and HLV ($HR_{LLV}$ and $HR_{HLV}$, respectively). The HR was estimated using any two consecutive HLV or LLV SCG events as follows

$$HR_j = 1 / (t_{SCG_{j+1}} - t_{SCG_j}) \qquad (1)$$

where $j$ is the index of SCG events in the LLV or HLV group. $t_{SCG_j}$ and $HR_j$ are the time of the $j^{th}$ SCG event peak and the estimated heart rate from the $j^{th}$ and $j+1^{th}$ SCG events. Equation 1 resulted in an inaccurate HR estimation when the two consecutive events were in two different LLV (or HLV) cycles such as the two LLV events encircled in Fig. 1. Although these LLV events are consecutive, they occur at the beginning and end of two different low lung volume cycles. Therefore, the time distance between them was not representative of a cardiac cycle duration. This longer time distance would result in a HR

TABLE I  OVERVIEW OF THE SUBJECTS' CHARACTERISTICS (MEAN ± SD).

| Age (years) | 29.4 ± 4.5 |
| Height (cm) | 173.1 ± 9.8 |
| Weight (kg) | 82.2 ± 18.3 |
| Number of subjects | 7 |

**Algorithm 1** Calculate heart rate during HLV and LLV

1: **Input:** SCG events, i (number of SCG events), respiratory flow rate signal, t (time vector)
2: **Output:** $HR_{LLV}$, $HR_{HLV}$
3: $LV$ = integral of respiratory flow rate
4: **for all** $i$ values **do**
5:    **if** $LV > 0$ **then**
6:       $SCG_{HLV}[j] \leftarrow SCG[i]$
7:       **for all** $SCG_{HLV}$ events **do**
8:          $HR[j] \leftarrow 1 / (t(SCG_{HLV}[j+1]) - t(SCG_{HLV}[j]))$
9:          **if** $HR[j] < 50$ **then**
10:            ignore $HR[j]$
11:          **Else**
12:            $HR_{HLV}[j] \leftarrow HR[j]$
13:          **end if**
14:       **end for**
15:    **else if** $LV < 0$ **then**
16:       $SCG_{LLV}[k] \leftarrow SCG[i]$
17:       **for all** $SCG_{HLV}$ events **do**
18:          $HR[k] \leftarrow 1 / (t(SCG_{LLV}[k+1]) - t(SCG_{LLV}[k]))$
19:          **if** $HR[k] < 50$ **then**
20:            ignore $HR[k]$
21:          **Else**
22:            $HR_{HLV}[k] \leftarrow HR[k]$
23:          **end if**
24:       **end for**
25:    **end if**
26: **end for**

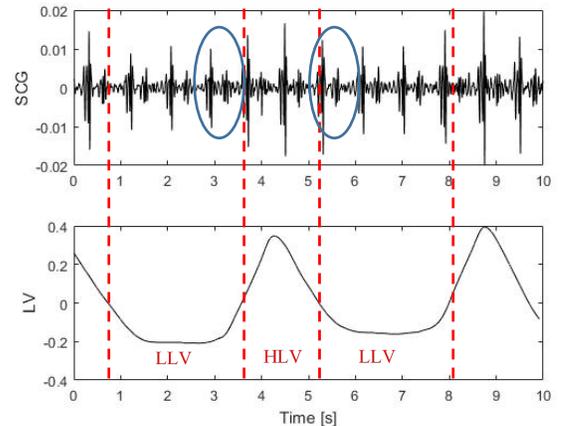

Fig. 1. (top) A 10-seconds portion of the SCG signal [Volts], (bottom) The lung volume signal [Volts] calculated from the integral of respiratory flow rate that was measured simultaneously with the SCG signal. The LLV and HLV portions of the lung volume signal are labeled.

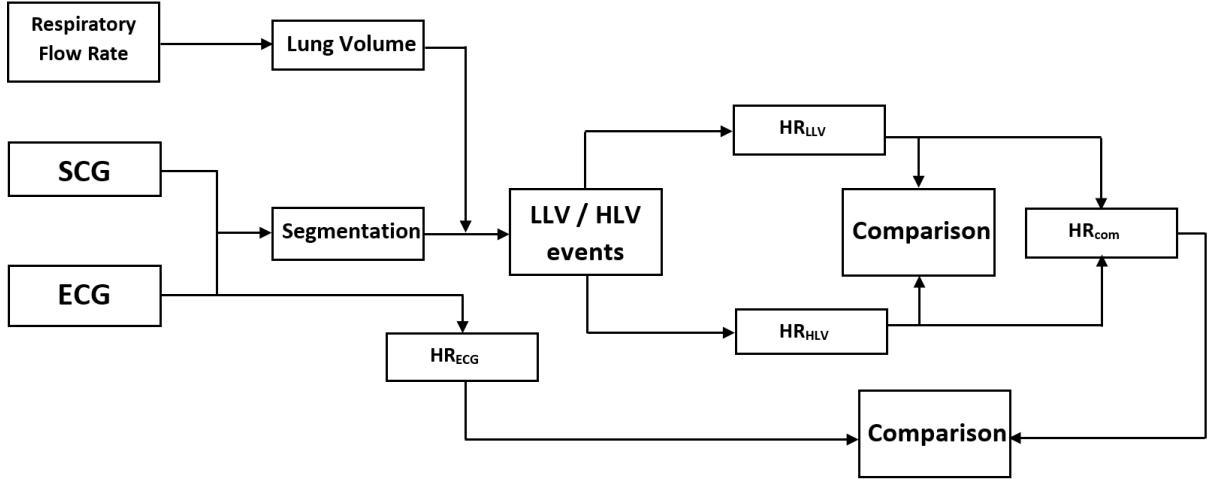

Fig. 2. Flowchart of the study.

TABLE II   ESTIMATED HEART RATE (BEAT PER MINUTE) USING HLV AND LLV SCG EVENTS DURING HIGH AND LOW LUNG VOLUME IN 3 DIFFERENT POSTURAL POSITIONS. THE VALUES ARE SHOWN AS MEAN ± SD.

| Subject # | Supine | | 45 degree - 1 | | 45 degree - 2 | | Sitting | |
|---|---|---|---|---|---|---|---|---|
| | LLV | HLV | LLV | HLV | LLV | HLV | LLV | HLV |
| 1 | 55.9 ± 2.9 | 69.0 ± 1.8 | 63.3 ± 3.4 | 72.2 ± 6.2 | 63.0 ± 4.0 | 70.7 ± 4.2 | 70.7 ± 3.7 | 74.4 ± 1.2 |
| 2 | 65.9 ± 0.9 | 66.6 ± 1.0 | 65.3 ± 3.1 | 66.1 ± 3.8 | 61.2 ± 2.7 | 63.9 ± 1.9 | 63.6 ± 1.5 | 66.0 ± 2.2 |
| 3 | 75.2 ± 1.8 | 80.3 ± 1.6 | 74.5 ± 4.0 | 78.7 ± 4.2 | 71.8 ± 4.5 | 75.9 ± 4.2 | 71.5 ± 2.2 | 81.3 ± 3.1 |
| 4 | 65.2 ± 3.3 | 76.4 ± 3.3 | 60.4 ± 4.4 | 70.0 ± 5.0 | 59.4 ± 4.4 | 68.5 ± 5.0 | 66.9 ± 4.1 | 75.5 ± 3.3 |
| 5 | 76.9 ± 5.9 | 82.4 ± 6.6 | 73.0 ± 3.9 | 77.4 ± 3.7 | 67.7 ± 3.6 | 73.5 ± 3.8 | 66.6 ± 2.2 | 72.6 ± 2.0 |
| 6 | 74.8 ± 4.8 | 83.2 ± 4.2 | 73.6 ± 4.3 | 78.2 ± 4.2 | 70.7 ± 6.1 | 75.2 ± 5.5 | 68.1 ± 3.7 | 73.4 ± 3.5 |
| 7 | 71.9 ± 2.4 | 78.3 ± 2.0 | 70.7 ± 4.1 | 78.3 ± 4.5 | 72.0 ± 4.2 | 79.2 ± 4.2 | 72.7 ± 4.7 | 80.8 ± 5.8 |

TABLE III   THE COMBINED HEART RATE (BEAT PER MINUTE) USING EQUATION 2.

| Subject # | Supine | 45 degree Trial 1 | 45 degree Trial 2 | Sitting |
|---|---|---|---|---|
| 1 | 62.2 | 67.9 | 66.3 | 72.1 |
| 2 | 66.2 | 65.6 | 62.1 | 64.3 |
| 3 | 77.2 | 76.1 | 73.1 | 75.3 |
| 4 | 69.4 | 64.0 | 63.8 | 70.6 |
| 5 | 79.1 | 75.2 | 70.7 | 69.6 |
| 6 | 78.6 | 75.9 | 73.1 | 70.9 |
| 7 | 74.7 | 73.3 | 75.0 | 75.3 |

estimation that was lower than the actual HR. Hence to avoid this, a HR minimum threshold of 50 bpm was chosen (as a rough preliminary threshold estimation). All the estimated HR values that were lower than this threshold were eliminated. Algorithm 1 shows a pseudo-code for the proposed heart rate monitoring algorithm.

The subject combined HR was also estimated from the $HR_{LLV}$ and $HR_{HLV}$ as follows

$$HR = \left(\sum_{i=1}^{m} HR_{LLV_i} + \sum_{j=1}^{n} HR_{HLV_j}\right)/(m + n) \quad (2)$$

where m and n are the total number of LLV and HLV events, respectively. The ECG signal was used as a gold standard in this study. The performance of the proposed SCG HR monitoring algorithm was tested against the standard ECG measurements. Fig. 2 shows the study flowchart.

## III. RESULTS

Table II shows the estimated heart rate using HLV and LLV SCG events during high and low lung volumes in the 3 different postural positions considered in this study. Results showed that all subjects had a higher HR during HLV compared to LLV in all postural positions. The combined HR estimations were listed in Table III. Fig. 3 compared the estimated LLV, HLV and combined HR at different postural positions for all the subjects.

## IV. DISCUSSIONS

### A. HR Algorithm Performance

Agreement between the ECG and SCG HR estimations was assessed using Bland-Altman analysis (Fig. 4). In this plot, the solid line represents the mean value of differences between ECG and SCG HR estimations, while the dashed lines show the 95% confidence interval (mean ± 1.96 SD). These results suggest general agreement between the proposed algorithm and the standard ECG method.

### B. Heart Rate Variability

Table IV shows the ratio between estimated HR during HLV and LLV ($HR_{HLV}/HR_{LLV}$) for all the subjects. For all the subjects and all the postural position, the ratio was larger than 1 indicating that subjects had a higher HR during HLV compared to LLV. The $HR_{HLV}/HR_{LLV}$ ratio was 1.11±0.07, 1.08±0.05, 1.09±0.04, and 1.09±0.04 for supine, 45 degree-first trial, 45 degree-second trial, and sitting positions, respectively. This can be due to the known physiological phenomenon of Respiratory Sinus Arrhythmia (RSA) [18]. RSA is an independent reflex, the exact explanation of which is not completely known. Although RSA was once explained as a secondary response of other known reflexes, the current understanding is that RSA is a separate physiological reflex [19]. A plausible justification for the evolution of RSA is related to ventilation-perfusion matching in the lungs. At high volume, the lungs are full of air (high ventilation) and an increase in heart rate will increase the amount of blood flow to the lungs (high perfusion). The opposite response is observed at low lung volume. A more well-known example of ventilation-perfusion matching is hypoxic pulmonary vasoconstriction, a reflex where blood vessels in the lungs constrict in response to low oxygen levels. This reflex reduces perfusion to areas of the lungs that have low ventilation, such as an area filled with fluid from pneumonia. The benefit of ventilation-perfusion matching is to ensure that all blood leaving the lungs has high $O_2$ saturation.

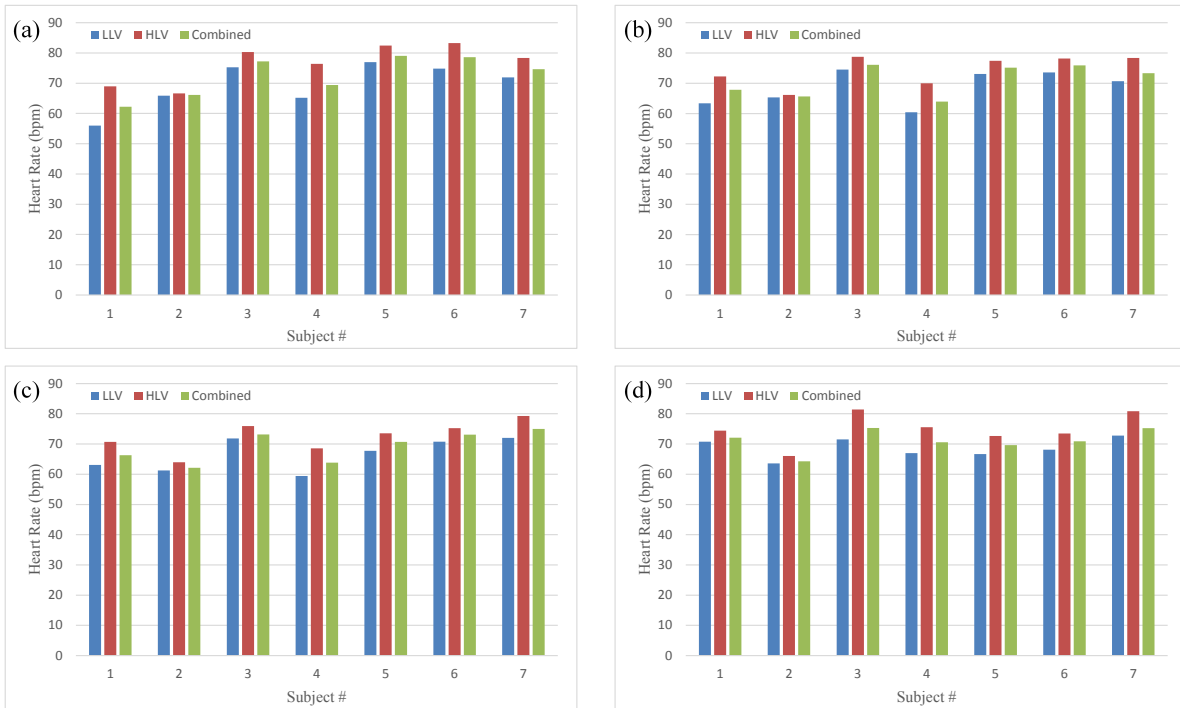

Fig. 3. The estimated LLV, HLV, and combined heart rate in (a) supine position, (b) 45-degree position first trial, (c) 45-degree position second trial, and (d) sitting position.

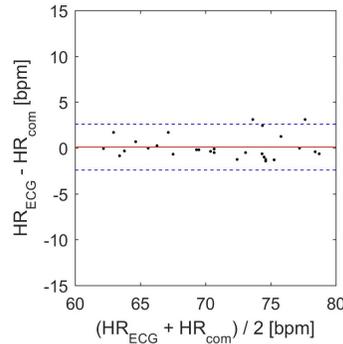

Fig. 4. Bland-Altman plot for $HR_{ECG}$ and $HR_{com}$. The solid line represents the mean (bias) difference between the HR values. The dashed lines show the 95% confidence interval.

### C. Application in Cardiovascular Diagnosis

Cardiovascular disease is a leading cause of death in the world. Analysis of blood flow dynamics [20]–[26] and heart related signals might result in improving current diagnostic methods, and eventually decreasing the mortality rate associated with cardiovascular disease. Cardiac time intervals (CTIs) have been widely used as clinical metrics for cardiovascular diagnosis. Results of the current study showed that the HR varies between different phases of the lung volume. This indicates that the cardiac cycles duration changes with the LV, which can lead to varying CTIs for different phases of the LV. Therefore, categorizing SCG cycles into two groups of LLV and HLV that contain SCG cycles with similar morphology can allow more accurate estimation of CTIs, and better signal characterization, and classification.

TABLE IV  THE RATIO BETWEEN ESTIMATED HEART RATE DURING HIGH AND LOW LUNG VOLUME $HR_{HLV}/HR_{LLV}$.

| Subject # | Supine | 45 degree - 1 | 45 degree - 2 | Sitting |
|---|---|---|---|---|
| 1 | 1.232449 | 1.140859 | 1.121364 | 1.051743 |
| 2 | 1.011316 | 1.011928 | 1.044089 | 1.038266 |
| 3 | 1.066846 | 1.05692 | 1.056712 | 1.13778 |
| 4 | 1.171733 | 1.158869 | 1.15357 | 1.128024 |
| 5 | 1.071803 | 1.059745 | 1.085252 | 1.089598 |
| 6 | 1.112487 | 1.062655 | 1.063825 | 1.078311 |
| 7 | 1.088873 | 1.108319 | 1.101045 | 1.111293 |

*D. Limitations*

The primary limitation of the study is the small number of participants. Narrow range of subject age, gender, weight, race, and clinical status might have affected results. Therefore, future studies need to enroll a larger number of subjects and cover more diverse population.

V. CONCLUSIONS

In this study, an ECG-independent algorithm for heart rate monitoring was proposed. This algorithm was used to estimate the heart rate of a control group during different phases of the lung volume. The proposed algorithm performance was tested against the standard ECG measurements. Results showed good agreement between the proposed and standard method. In addition, the new method was able to detect HR changes with respiration.